\documentclass[twocolumn,prl,superscriptaddress,preprintnumbers]{revtex4}[10pt]
\pdfoutput=1
\usepackage{amsmath,amssymb,graphicx} 
\usepackage{epsf,verbatim}
\usepackage{pstricks}
\usepackage{hyperref}
\usepackage{comment}
\usepackage{cancel}

\newcommand{\squishlist}{
 \begin{list}{$\bullet$}
  { \setlength{\itemsep}{0pt}
     \setlength{\parsep}{3pt}
     \setlength{\topsep}{3pt}
     \setlength{\partopsep}{0pt}
     \setlength{\leftmargin}{1.5em}
     \setlength{\labelwidth}{1em}
     \setlength{\labelsep}{0.5em} } }

\newcommand{\squishlisttwo}{
 \begin{list}{$\bullet$}
  { \setlength{\itemsep}{0pt}
     \setlength{\parsep}{0pt}
    \setlength{\topsep}{0pt}
    \setlength{\partopsep}{0pt}
    \setlength{\leftmargin}{2em}
    \setlength{\labelwidth}{1.5em}
    \setlength{\labelsep}{0.5em} } }

\newcommand{\squishend}{
  \end{list}  }
  
  \newcommand{\ww}{
$W^+W^-$
}

\newcommand{\ifb}{\mathrm{fb}^{-1}}

\newcommand{\pb}{\mathrm{pb}}

\newcommand{\gev}{~\mathrm{GeV}}

\newcommand{\wz}{$W^\pm Z$~}

\newcommand{\wh}{$W^\pm h$~}

\newcommand{\wg}{$W^\pm \gamma$~}

\newcommand{\hww}{$h\rightarrow W^+W^-$~}

\newcommand{\hgg}{$h\rightarrow \gamma\gamma$~}

\begin{document}

\preprint{YITP-SB-12-24}

\title{Charginos Hiding In Plain Sight }

\author{David Curtin}
\affiliation{C. N. Yang Institute for Theoretical Physics, Stony Brook University, 
 Stony Brook, NY 11794}

\author{Prerit Jaiswal}
\affiliation{C. N. Yang Institute for Theoretical Physics, Stony Brook University, 
 Stony Brook, NY 11794}
  \affiliation{Department of Physics, Brookhaven National Laboratory, Upton, NY 11973, USA }
 
 \author{Patrick Meade}
\affiliation{C. N. Yang Institute for Theoretical Physics, Stony Brook University, 
 Stony Brook, NY 11794}

\begin{abstract}
Recent 7 TeV 5/fb measurements by ATLAS and CMS have measured both overall and differential $W^+W^-$ cross sections that differ from NLO SM predictions. While these measurements aren't statistically significant enough to rule out the SM, we demonstrate that the data from both experiments can be better fit with the inclusion of electroweak gauginos with masses of $\mathcal{O}(100)$ GeV.   We show that these new states are consistent with other experimental searches/measurements and can have ramifications for Higgs phenomenology. Additionally, we show how the first measurements of the \ww cross section at 8 TeV by CMS strengthen our conclusions.
\end{abstract}

\maketitle

 \setcounter{equation}{0} \setcounter{footnote}{0}

\section{Introduction}
\label{s.intro} \setcounter{equation}{0}
In~\cite{wwatlas5,wwcms5} ATLAS and CMS have measured the \ww cross section using the full $\sim 5/$fb LHC7 dataset.  ATLAS measured a cross section of  $53.4 \pm 2.1(\mathrm{stat.}) \pm 4.5(\mathrm{syst.}) \pm 2.1 (\mathrm{lumi.})$~pb compared with a NLO theory prediction of $45.1\pm 2.8$ pb, while CMS found a cross section of $52.4\pm2.0 (\mathrm{stat.}) \pm 4.5 (\mathrm{syst.}) \pm 1.2 (\mathrm{lumi.})$~pb compared to a NLO theory prediction of $47.0 \pm 2.0$~pb.\footnote{The difference between the ATLAS and CMS NLO predictions come from ATLAS using NLO MC while CMS took the cross section from a theory paper.}  Both of these results are consistent within the error bars, but they are also both high and more consistent with each other than with the SM.   Even more recently, CMS released a measurement of the SM \ww cross section at 8 TeV~\cite{wwcms8tev} that was also higher than the SM cross section and even more discrepant with the SM than the 7 TeV measurement~\cite{wwcms5}.  Understanding the \ww cross section is crucial for Higgs searches as well as any new physics search containing more than one lepton.   In this letter, we investigate the consequences for new physics that could be responsible for the enhanced values of the measured \ww cross sections.

In \cite{wwatlas5,wwcms5,wwcms8tev} the \ww cross section was measured in the fully leptonic final state.  These analyses were designed to measure the SM, not to exclude new physics, which is reflected in their rather inclusive cuts. Therefore, new physics that produces OS leptons and MET could be present in these measurements.  Based on the differential distributions in~\cite{wwatlas5,wwcms5,wwcms8tev},  new physics that contaminates these measurements would have to have kinematics at least similar to SM \ww events, otherwise there would be an obvious discrepancy.  While it is quite probable that the discrepancies in the total cross section and differential distributions are due to insufficient background modeling, we demonstrate that SM NLO \ww combined with the inclusion of new EW processes fits the data significantly better than the current SM prediction alone.

\begin{figure*}[t] 
    \includegraphics[width=6in]{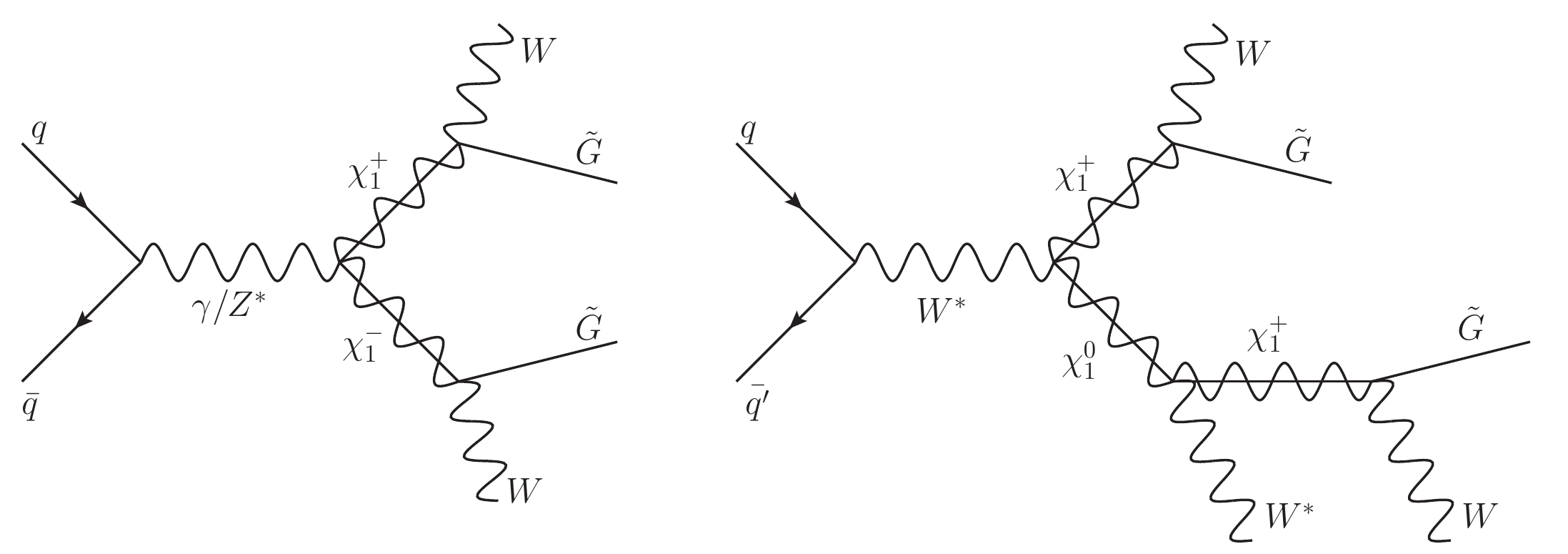}
   \caption{Examples of electroweak gaugino production and decay for our gauge-mediated SUSY benchmark model (Chargino pairs on the left and Chargino-Neutralino on the right). Both processes give a \ww + MET final state, since the decay products of the off-shell $W^*$ in the right diagram are typically too soft to be detected. }   \label{fig:prod}
\end{figure*}

Supersymmetry provides an example of new physics that could significantly affect EW cross section measurements.  In particular, light EWinos provide a production and decay mechanism to generate multigauge boson final states as shown in Fig.~\ref{fig:prod} for a gauge-mediated scenario.

While colored particles typically have bounds of $\mathcal{O}(\mathrm{TeV})$, Charginos can in principle be as light as $\mathcal{O}(100)\gev$~\cite{charginoLEPbound}, and even massless neutralinos are compatible with all collider constraints~\cite{lightneutralino}.  For EWinos with masses $\mathcal{O}(100)\gev$, the production cross section is $\mathcal{O}(1-10)$ pb for $\chi^+\chi^-$ and $\chi^\pm \chi^0$ at the LHC, precisely the range of interest for explaining the \ww cross section discrepancy from the NLO SM prediction.  Accounting for the \ww discrepancy fixes the cross section and mass scale, which turns out to automatically dictate that the kinematics of onshell W's from $\chi^+\chi^-$ decays are very similar to the SM.  In this letter we will demonstrate that $\chi^2/N_{bins}$ of the \ww measurement is significantly better for all ATLAS and CMS distributions when EWinos are included.  In addition to \ww production, \wz, \wh and \wg + MET final states are generically also produced via $\chi^\pm\chi^0$ production.  However, current bounds on the \wz and \wh final state are quite strong~\cite{atlastrinew,atlashbb}.  We show that there are several classes of models where the \ww discrepancy can be accounted for without being in conflict with existing experimental searches.

Weak scale gauginos not only affect SM multi-gauge boson measurements but also can effect Higgs phenomenology and measurements.  If new physics were to contaminate the signal and control regions of the \hww search, a Higgs discovery~\cite{atlashobs,cmshobs} in this channel could be affected~\cite{dieter}.  Given the similar kinematics to SM \ww we find that the presence of EWinos would only be manifest as an upscaling of the SM \ww background in the control region to match the data.  The presence of very light EWinos could also create new Higgs production channels and modify loop induced decay processes.  For example, new contributions to $W^\pm h$ production can be realized with EWinos.  For decays, EWinos themselves can modify the \hgg rate in a number of ways~\cite{EwinoHiggs, EwinoHiggsReview}, however this is also correlated to the production cross section for EWinos themselves.  Without specifying the mechanism of generating a 125 GeV Higgs in SUSY, which is not necessarily coupled to the \ww cross section, it is impossible to make definite predictions for changes in Higgs phenomenology.

In the rest of this letter, we will quantitatively demonstrate the effects of a particular SUSY scenario for the \ww measurement at 7 TeV and 8 TeV.  We then investigate the bounds on these scenarios, and their contributions to other multi-gauge boson and Higgs measurements/searches.  Finally we discuss the impact of this scenario and possible ways to test for it and other closely related scenarios in the future.  While the discrepancies in \ww may simply be due to background modeling, this letter clearly demonstrates that EW charginos could have been hiding in plain sight, and can improve a number of SM measurements done thus far at the LHC.

\section{\ww cross section at 7 TeV}
\label{wwxsec7} \setcounter{equation}{0}

\begin{figure*}[t] 
   \centering
   \vspace{-6mm}
 \begin{tabular}{c}
    \includegraphics[width=7in]{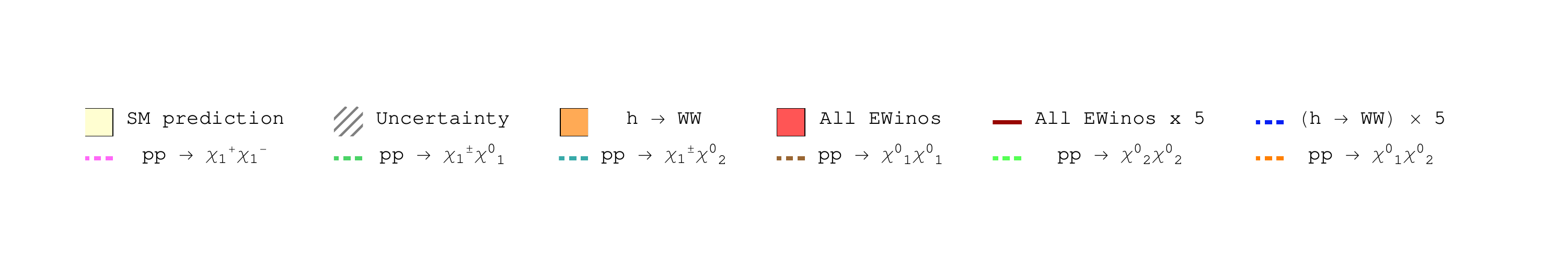} \vspace*{-9mm}
    \\
   \begin{tabular}{cc}
   \includegraphics[width=3in]
{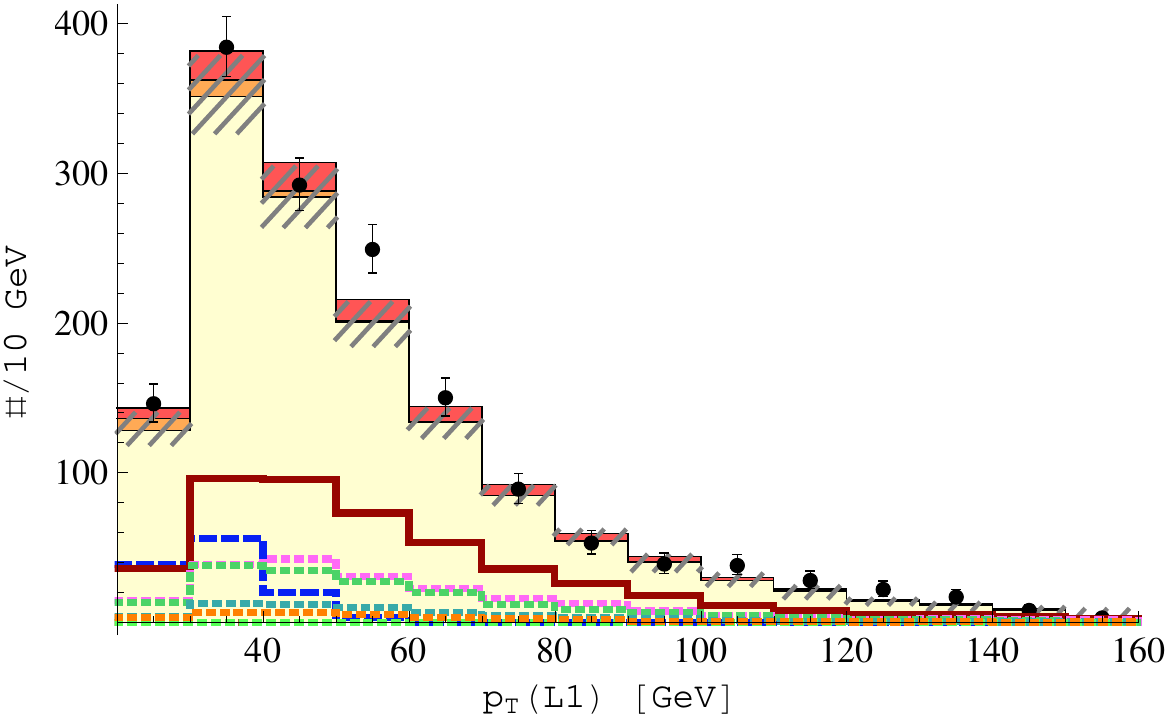} 
& \includegraphics[width=3in]
{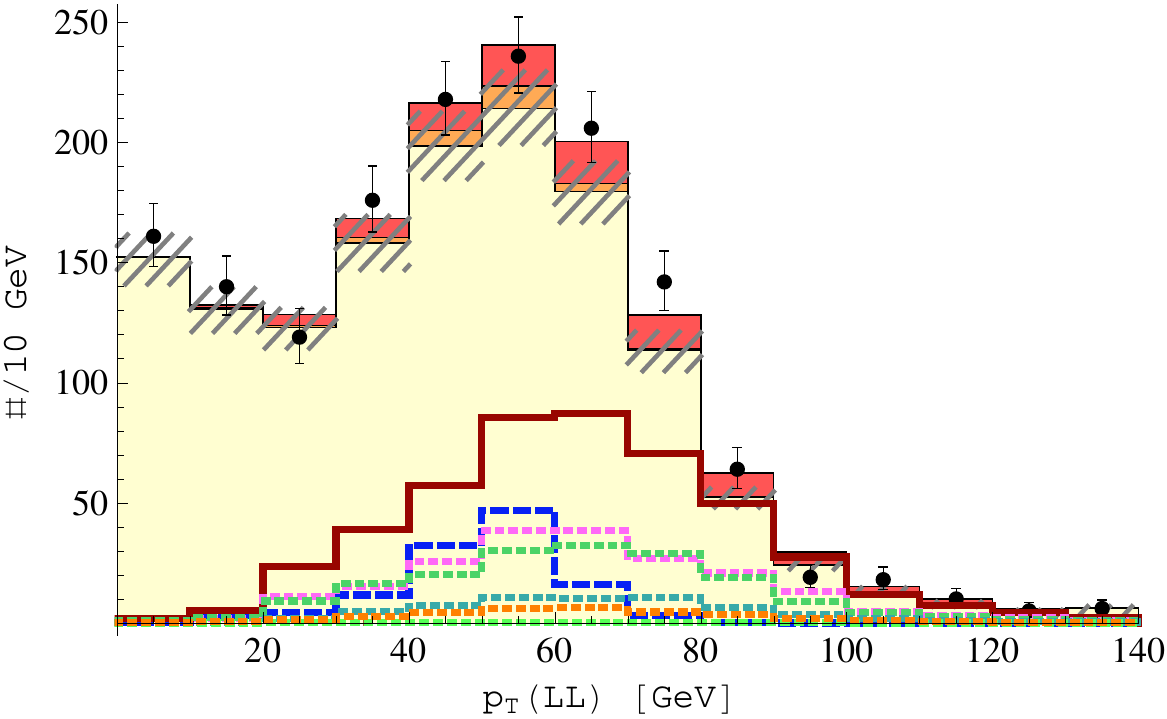}
\\
      \includegraphics[width=3in]
      {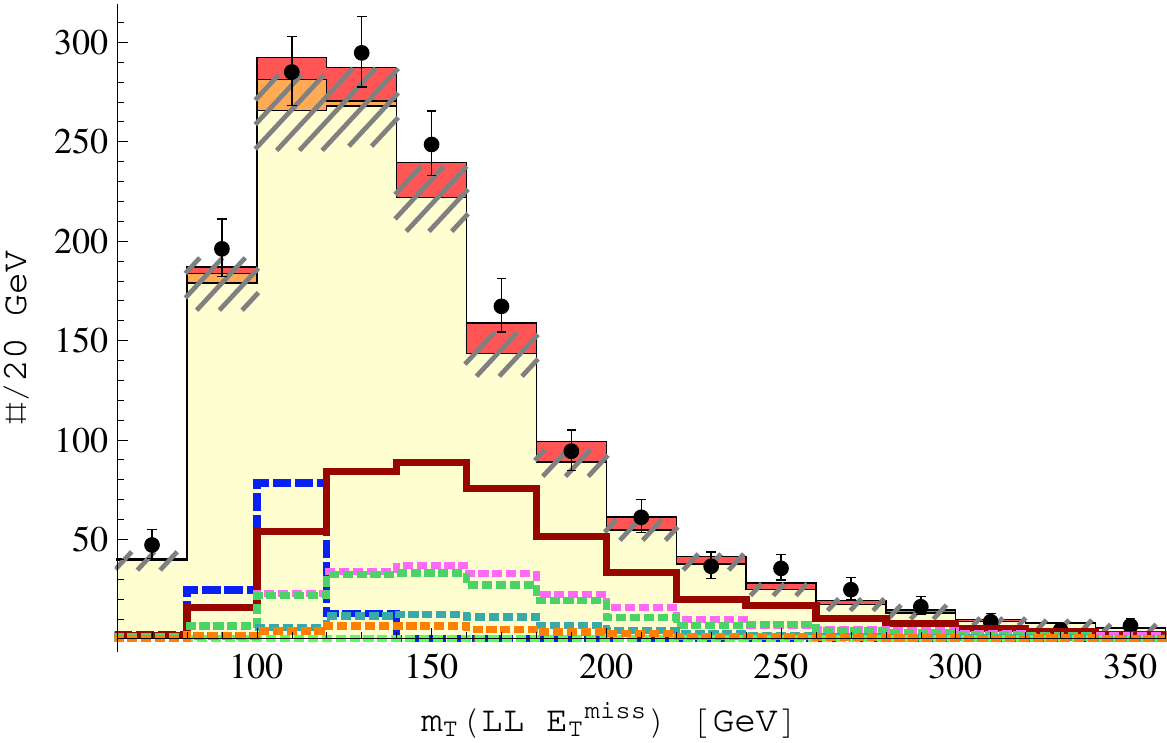}
       & 
       \includegraphics[width=3in]
       {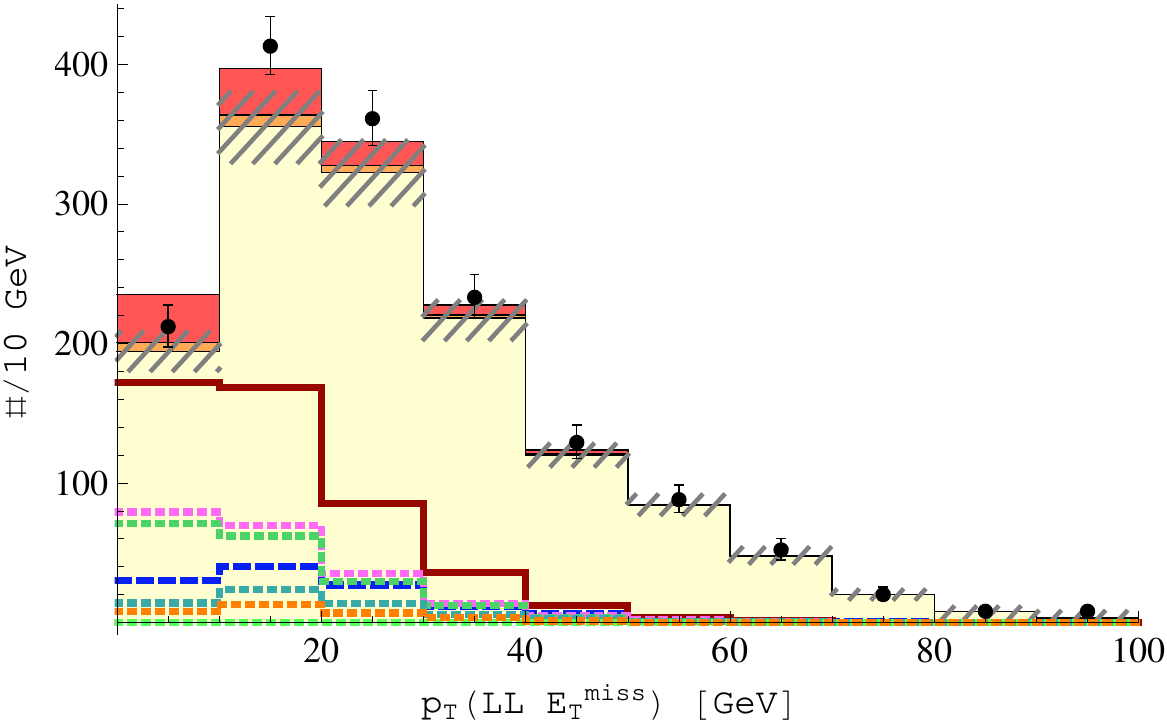}
       
   \end{tabular}

 \end{tabular}
   
   \caption{
   The total SM prediction (signal + background) from the LHC7 ATLAS \ww study \cite{wwatlas5}, with additional contributions from a $125 \gev$ SM higgs and chargino pair production in the best-fit gauge mediated scenario $m_{\tilde \chi^+_1} = 110 \gev$  shown. The gray hashed bands represent the uncertainty of the SM prediction.
   }
   \label{fig2}
\end{figure*}

ATLAS \cite{wwatlas5} and CMS \cite{wwcms5} measure the \ww production cross section in the dileptonic final state $ee$, $\mu\mu$ or $e \mu$ with $5~ \ifb$ of LHC7 data. The main backgrounds to $pp \rightarrow W^- W^- \rightarrow \ell^+ \ell^- \nu \bar \nu$ are Drell-Yan, top quark, $W$ + jet and other diboson production.   ATLAS imposes a series of cuts designed to remove excess jet activity and focus on real OS leptons (not from a Z) + MET, without an upper cut on MET.  CMS imposes similar if not softer cuts, but has different restrictions on the dilepton system overall and imposes additional vetoes, resulting in higher signal purity. Both analyses have an acceptance of about 6\% for pair-produced $W$'s in the fully leptonic channel. ATLAS and CMS also use different methods to estimate their acceptances for signal. In the end their similar but still different approaches result in extremely consistent measured central values for the \ww cross section, perhaps making the particular value measured quite compelling.

To demonstrate the agreement or lack thereof between data and the SM, kinematic distributions from ATLAS are shown in Figure~\ref{fig2} (CMS has similar but slightly fewer kinematic distributions available). There is some disagreement, not only in the overall normalization but also in the shape -- bins at high and low values of the kinematic variables generally fit quite well, while the middle bins display somewhat more significant excesses. As mentioned earlier, if new particles are produced which then decay into OS leptons and missing energy, one could potentially explain discrepancies with the data. Within the MSSM framework, pair-produced charginos are a natural candidate for such particles, though our statements are more broadly applicable in the simplified model context.

In order to display similar kinematics to SM \ww and improve agreement with data, the simplest possibility is for charginos to decay via on-shell $W$'s with a production cross section of a few $\pb$, setting a rough upper bound on their mass scale. Slightly more complicated possibilities arise through decays via either off shell W's or slepton decays.
Taking into account the chargino mass bound from LEP \cite{charginoLEPbound}, this implies $100 \gev \lesssim m_{\tilde \chi^\pm_1} \lesssim 130 \gev$, wino-like charginos, and a mass gap to an invisible detector-stable particle larger than $m_W$\footnote{If the neutral lightest state decays within the detector to soft particles, such as in RPV this potentially could also fit the data equally well}. This can easily be achieved both in gravity mediation (with a light bino LSP) or gauge mediation (with a gravitino LSP).  However, recent trilepton searches from ATLAS \cite{atlastrinew}, and searches for associated production of \wh in the $b\bar{b}$ channel~\cite{atlashbb}, significantly constrain $\chi^\pm\chi^0$ decays into \wh or \wz final states.  We will discuss these bounds later in this letter, but ultimately they lead to two possible SUSY scenarios for increasing the \ww cross section that remain in agreement with all other experimental data.  The first is a gauge mediated scenario with chargino NLSP, resulting in exclusively \ww + MET final states.  The second scenario, which is realized in gravity mediation, relies on an intermediate slepton to avoid $\chi_2^0\rightarrow \chi_1^0 h/Z$ decays and soften lepton $p_T$'s sufficiently to avoid bounds.  In this letter we focus on the first scenario as a benchmark while the second, which doesn't rely on actual $W$'s to affect the \ww cross section, will be described in more detail in~\cite{workinprogress}.

The benchmark point we use as an example is a gauge mediation inspired spectrum with a chargino NLSP, where low $\tan \beta$ and a high higgsino fraction makes the two lightest neutralinos heavier than the chargino \cite{Kribs:2008hq}. For our demonstration we chose $m_{\chi^\pm_1} \approx 110 \gev$, $m_{\chi^0_1} \approx 113 \gev$ and $m_{\chi^0_2} \approx 130 \gev$.  The most important parameter is the chargino mass, since it determines the $\chi^\pm_1 \chi^\mp_1$ and $\chi^0_1 \chi^\pm_1$ pair production cross sections.

For our example point, the NLO pair production cross section (calculated in \texttt{Prospino} \cite{prospino})  is 4.3 pb.  The large cross section comes from the sum of all Chagino/Neutralino mode production modes, since in the GMSB scenario all states decay to $W^+W^-$, and offsets the smaller direct production cross section for higgsino-rich chargino pairs compared to winos. The additional decay products from neutralinos decaying to charginos are typically too soft to affect any searches, see Fig. \ref{fig:prod}.  To estimate the chargino's effect on the \ww distributions, we generated $p p \rightarrow \chi \chi  \rightarrow \ell^+ \ell^- \tilde G \tilde G + X$ events in \texttt{Pythia~8} \cite{pythia8}, interfaced with \texttt{Pythia~6.4} \cite{pythia64} for the hard process. ($\ell = e/\mu/\tau$, $\tilde G$ is the practically massless gravitino, and $X$ are the soft particles from the decay of a possible off-shell $W$.) The events are passed to a \texttt{FastJet~3.0.2} \cite{fastjet}  based code that performs the same series of event reconstruction  and cut steps as the respective \ww cross section measurement analyses. This includes a rudimentary detector simulation that models geometric acceptances, jet reconstruction, and imposes lepton and photon isolation requirements and detector efficiencies, according to the ATLAS/CMS  specifications. 
	
The combined acceptance of dileptonic EWino events is about 4\% for the ATLAS analysis and 2.5\% for the CMS analysis, which imposes an additional $p_T^{\ell \ell}$ cut.  These figures are comparable to the quoted acceptances for dileptonic \ww events, which is expected given the $W$-like kinematics of the chargino decay and makes it plausible that the few-pb of chargino pair production makes up the few-pb-excess seen in the \ww cross section measurements. 

Figure~\ref{fig2} shows the chargino contributions stacked on top of the SM expectation for our example point. (We have also included the effect of a 125 GeV SM higgs decaying to $W^+W^-$, which is a small but non-negligible effect.) By eye it is clear that the agreement with data is very much improved in \emph{all kinematic distributions} (including two that are not shown): the charginos preferentially fill in the bins where the data disagrees most with the SM prediction, while leaving those bins where the SM agrees well with data relatively unaffected. The same can be said of the CMS distributions. Including charginos improves the $\chi^2/N_{bins}$ from $\sim 1$ for SM alone to $\lesssim 0.5$ in all ATLAS distributions,  quantifying the above statements.  For CMS, the reduction in $\chi^2/N_{bins}$ is not quite as significant, but the agreement with data is still improved in all distributions (except $p_{T(L_1)}$, where the level of agreement is unchanged).

\section{\ww cross section at 8 TeV}
\label{wwxsec8} \setcounter{equation}{0}

\begin{figure*}[t] 
   \centering
   \vspace{-6mm}
 \begin{tabular}{c}
   \begin{tabular}{cc}
   \includegraphics[width=3in]
{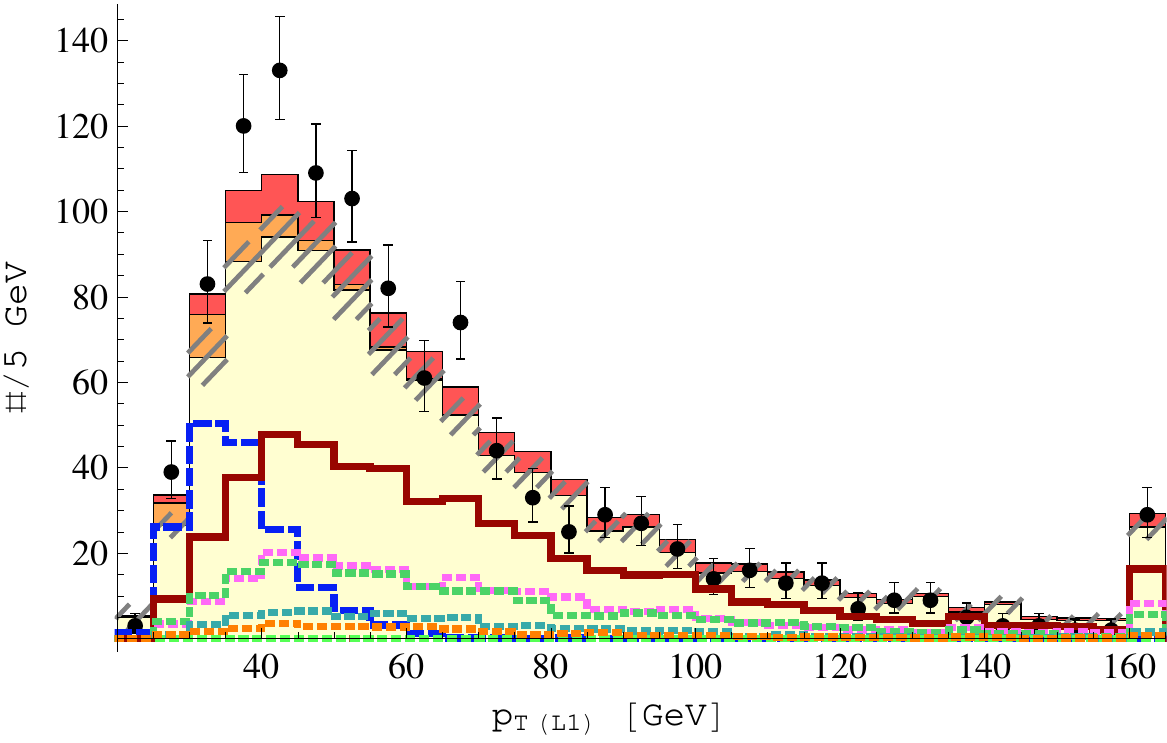} 
& \includegraphics[width=3in]
{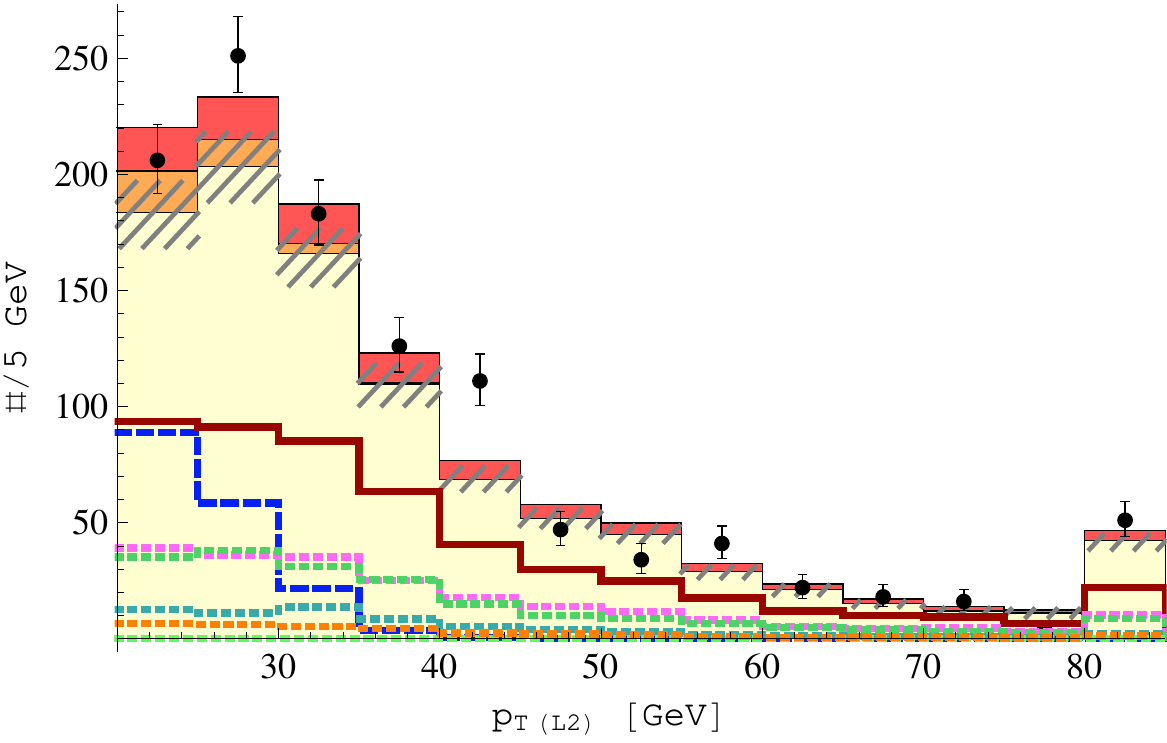}
\\
      \includegraphics[width=3in]
      {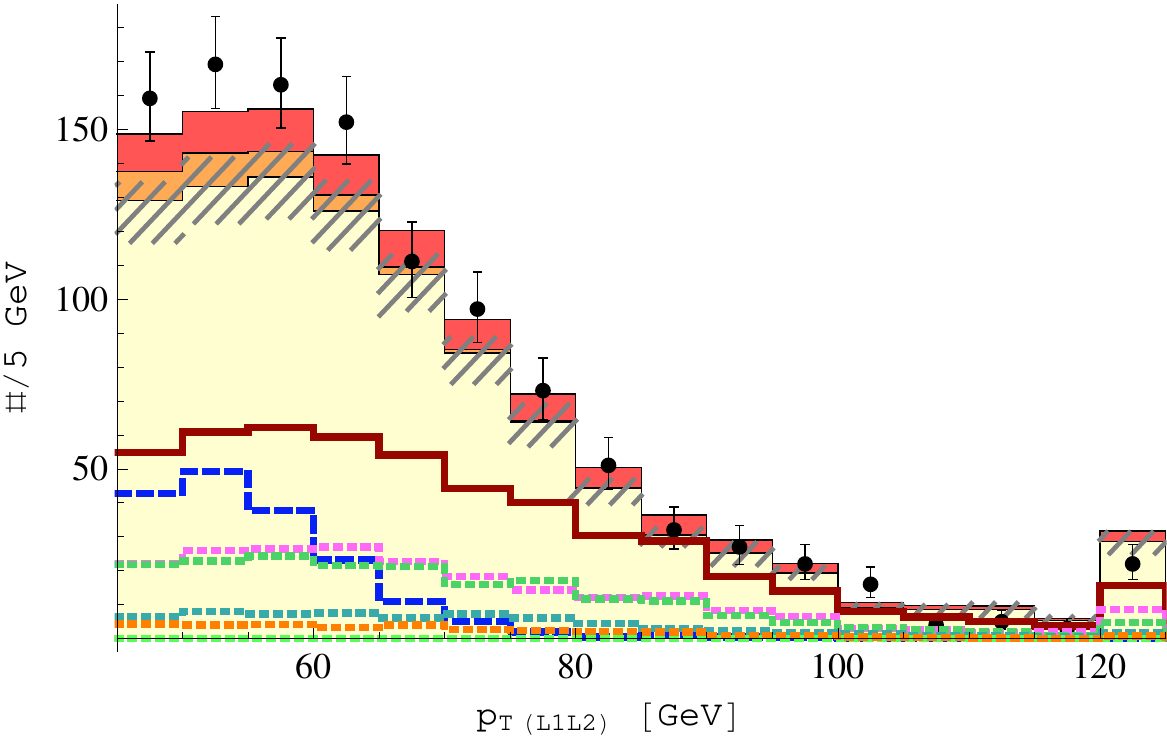}
       & 
       \includegraphics[width=3in]
       {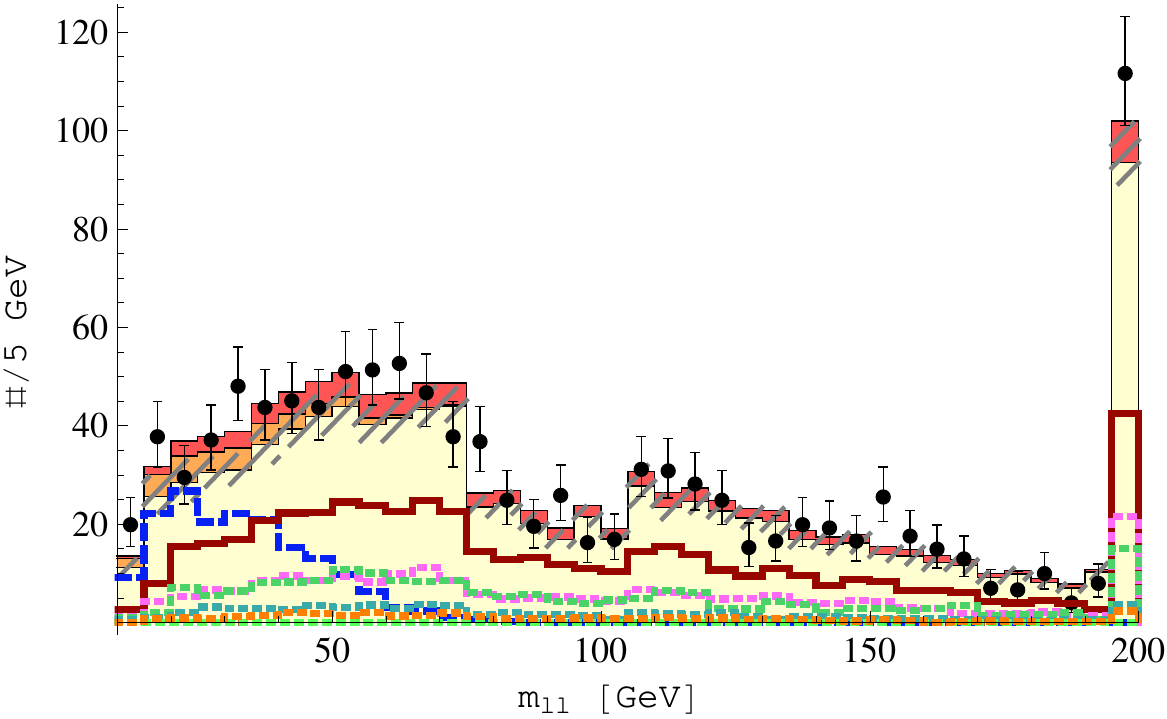}
       
   \end{tabular}

 \end{tabular}
   
   \caption{
   The total SM prediction (signal + background) from the LHC8 CMS \ww study \cite{wwcms8tev}, with additional contributions from a $125 \gev$ SM higgs and chargino pair production in the best-fit chargino NLSP scenario ($m_{\tilde \chi^+_1}= 110 \gev$)  shown. The gray hashed bands represent the uncertainty of the SM prediction. The legend is the same as for Figure~\ref{fig2}
   }
   \label{fig3}
\end{figure*}

Given that the 7 TeV measurements of ATLAS and CMS can not rule out the SM with their current level of precision, it is still possible that both ATLAS and CMS both measured an upwards fluctuation at 7 TeV.  However, if new physics was the cause of the discrepancy with the SM, then the \ww cross section at 8 TeV should also be high.  As mentioned in the introduction, this was recently shown to be the case by CMS~\cite{wwcms8tev}.  The LHC8 measured value from CMS is $69.9 \pm 2.8(\mathrm{stat.}) \pm 5.6(\mathrm{syst.}) \pm 3.1 (\mathrm{lumi.})$~pb with an NLO SM prediction of $57.3\scriptsize\begin{array}{c} +2.4\\ -1.6 \end{array}$pb.  Recently, the 8 TeV LHC NLO EW corrections were also computed~\cite{ewnlo}, and the effect on the total cross section from EW corrections is negative which would only increase the discrepancy.  

In Figure~\ref{fig3} we show the contribution from both charginos and the Higgs to the differential distributions from~\cite{wwcms8tev}. Since the cross section discrepancy has grown compared to the ATLAS LHC7 results, the agreement of the SM prediction with the data is quite poor, and computing $\chi^2/N_{bins}$ for all the plots yields a combined $p$-value of $\sim 10^{-3}$ (assuming approximately gaussian errors). Similar to LHC7 results, the inclusion of the charginos reduces the $\chi^2/N_{bins}$ by about a factor of 2, significantly improving the $p$-value of the combined fit to 0.3 for charginos + SM. The contribution of the Higgs to the \ww measurement is larger at 8 TeV than at 7 TeV, but is still subdominant to the charginos in improving the fit, and in the differential distributions it enters in different kinematic regimes than the chargino contribution. Adding only the higgs contributions improves the $p$-value to $\sim 0.1$, while adding both the higgs and chargino contributions yields a combined $p$-value of $\sim 0.75$, a very significant improvement.

\section{Constraints on Electroweakinos}
\label{charginobounds} \setcounter{equation}{0} 

For particles charged under the electroweak symmetries alone the LHC bounds are weaker than for strongly interacting particles. This is mostly due to the decreased production cross section, resulting in a lower mass reach, but also because the lower mass scale implies kinematic distributions more similar to the SM in many cases -- such as the one we are considering in this letter. Nevertheless, recent LHC searches \cite{cmsmulti, atlastrinew, atlasdinew, atlasdinewinc} are starting to approach the sensitivity necessary to exclude $\mathcal{O}(100 \gev)$ electroweakinos, and we have to examine the bounds carefully. 

In principle many SUSY scenarios could give rise to a shift in SM \ww measurements. Perhaps the most obvious example is gravity mediation with Wino-like charginos that have a mass of $\sim 110  - 130 \gev$ along with a light Bino-like neutralino LSP, and $m_{\chi^0_1} < (m_{\chi^\pm_1} - m_{W})$. However, $\chi^\pm_1 \chi^0_2$ associated production followed by a $\chi^0_2 \rightarrow Z \chi^0_1$ decay produces a trilepton signal that is completely excluded by recent ALTAS trilepton bounds \cite{atlastrinew}, regardless of whether the $Z$ is on- or off-shell. One could imagine diluting the trilepton signal by increasing $M_2$ to open up the $\chi^0_2 \rightarrow h \chi^0_1$ decay, but in this case the resulting $hW^\pm$ final state is excluded by $Wh \rightarrow Wb\bar{b}$ searches \cite{atlashbb}. 

Gravity-mediation could still provide a viable scenario if sleptons are lighter than the EWinos, in which case their pair production would contribute to the $\ell^+ \ell^-$ + MET final state, while the decay of EWinos to sleptons + soft leptons might ameliorate the trilepton signal. We will examine this scenario in more detail in \cite{workinprogress}. 

The benchmark scenario of the previous sections is realized within gauge mediation, where $M_1 > M_2$, low $\tan \beta$ and a large degree of wino-higgsino mixing can produce the \emph{Chargino-NLSP} spectrum we described. Before we examine the bounds on this scenario in detail there are two other choices of NLSP to consider: bino neutralino and wino neutralino. 

In the \emph{Bino-NLSP} scenario, the bino decay $\chi^0_1 \rightarrow \tilde G \gamma$ leaves a striking diphoton signature for SUSY production. As outlined in \cite{ggmLHC}, LHC diphoton + MET searches \cite{LHCgaga} exclude chargino pair production and decay for $m_{\tilde \chi^\pm_1} \lesssim 450 \gev$, making their pair production cross section too small to enhance the measured \ww cross section. Photon constraints can be partially avoided with a \emph{Wino-Like NLSP}: for $M_1 > M_2$ with low wino-higgsino mixing the neutralino-chargino mass-splitting is so small that $\tilde\chi^\pm_1$ decays directly to $W^\pm \tilde G$, making it a  chargino co-NLSP and avoiding stringent diphoton + MET bounds. $\chi^0_1 \chi^\pm_1$ associated production still produces some single-photon signal (depending on the branching ratios of $\chi^0_1 \rightarrow \tilde G \gamma$ vs $\tilde G Z$), and a CDF $\ell + \gamma$ search excludes $m_{\tilde \chi^\pm_1} \lesssim 135 \gev$ \cite{neutralinostevatron}. A chargino at this mass could somewhat ameliorate the discrepancy in the measured \ww cross section, but $\mathrm{Br}(\chi^0_1 \rightarrow Z \tilde G)$ is high enough for new ATLAS trilepton bounds \cite{atlastrinew} to completely exclude this scenario as well.

We are therefore left with the chargino-NLSP scenario as the only possible realization of gauge mediation to contribute significantly to the \ww final state. The trilepton signal is suppressed (though not completely absent, due to the non-negligible $\chi^\pm_1$-$\chi^0_2$ mass difference), but chargino-neutralino production produces a significant amount of \emph{same-sign dilepton} signal, making constraints from new ATLAS dilepton searches \cite{atlasdinew, atlasdinewinc} relevant. We simulated the signal produced by our scenario in these searches \cite{atlastrinew, atlasdinew, atlashbb, atlasdinewinc, atlastriold, atlaswwres} using the same Monte-Carlo setup as for the \ww cross section measurement. Each search is still consistent well within one sigma, though it could be possible for same-sign dilepton searches to discover this scenario with the full 8 TeV LHC data set.

\section{Gauge Boson Phenomenology}
\label{npgauge} \setcounter{equation}{0} 

Given the current bounds on trilepton \cite{atlastrinew} signatures, any new physics must primarily affect only the \ww cross section, leaving \wz and \wg mostly unaffected.  

To illustrate this, consider the gravity mediated scenario discussed in the previous section, with Winos always decaying to a Bino-like neutralino LSP via on-shell $W$'s and $Z$'s. In this case the trilepton bounds push the allowed mass of the Winos to $m_{\chi^\pm_1} \gtrsim 190 \gev$, which makes the wino pair production cross section so small that the $\chi^2/N_{dof}$ improvement of the \ww measurement is negligible, less than $\sim 5\%$.

Since our chargino NLSP scenario evades these trilepton bounds there is no affect on multi-gauge boson phenomenology other than multi-$W$.  There will be signatures of same-sign $W$ gauge boson production with additional soft jets or leptons arising from $\chi^\pm\chi^0$ production and decay.  As discussed in the previous section, same-sign dilepton searches \cite{atlasdinew, atlasdinewinc} are not yet sensitive enough to rule out this signal. 

The possibly viable gravity mediated scenario with intermediate sleptons \cite{workinprogress} could feature additional ``gauge boson'' signatures, because in addition to producing $\ell \ell + \mathrm{MET}$ final states (even though no $W$'s are involved) there is also the possibility for $\ell + \mathrm{MET}$ production, showing up in single $W$-measurements.

\section{Higgs Phenomenology}
\label{nphiggs} \setcounter{equation}{0}

Modifying the effective \ww cross section through BSM contributions could significantly affect \hww measurements since both ATLAS \cite{atlasHww} and CMS \cite{cmsHww} searches use data-driven techniques to estimate  \ww background. The Monte Carlo output is normalized to fit the data in a control region, and that ``renormalization" is carried over into the signal region. 

However, we find that generically the Higgs search sensitivities are not modified. In a BSM scenario like ours, where the kinematics are  very similar to $W^+W^-$, the control and signal regions are contaminated in proportion to $W^+W^-$, and the upscaling of the \ww contribution in the control region correctly `predicts' the BSM background in the signal region. On the other hand, the examples studied by \cite{dieter}, which preferentially contaminate the control region due to their kinematics being very different from $W^+W^-$, are excluded by trilepton searches \cite{CMStrilepton}. It would be interesting to see if there was a scenario that did not fall into either of these categories, i.e. contaminating preferentially the control region without generating a large trilepton signal. 

Light EWinos can also generate Higgs production that mimics associated \wh production to the point that it becomes a constraint, as shown previously.  In principle there could be some contribution to \wh, but as an example we can look at a gravity mediated scenario with a Bino-like Neutralino LSP and a Higgsino-like NLSP, such that $\chi^\pm\chi^0_2$ always leads a final state of $W^\pm h \chi^0\chi^0$.  To avoid constraints from existing \wh measurements requires a mass scale of $m_{\chi^\pm}\gtrsim 160\;\mathrm{GeV}$.  However, the high Higgsino fraction in this case lowers the cross section for $\chi^+\chi^-$ and the improvement of the $\chi^2$ for \ww is $\lesssim 5\%$.  Therefore similar to \wz there is typically no signature in \wh when an appreciable effect is measurable in the \ww cross section.

As a final avenue, light EWinos can also modify loop induced decay processes of the Higgs, or in principle provide new decay modes for the Higgs.  In our particular benchmark scenario this is manifested as a 15\% increase in the partial width of \hgg.  This is of course tantalizing given the fact that both ATLAS~\cite{atlashobs} and CMS~\cite{cmshobs} observed an increased rate of \hgg compared to the SM in their discoveries of a Higgs like state.  However, within the context of the SUSY the Higgs mass is tightly intertwined with the resulting phenomenology more so than in other models, thus without specifying how to generate a 125 GeV Higgs there could be other much larger contributions that affect Higgs production and decay to the point where a generic prediction based solely on accounting for the \ww cross section is impossible.  It is certainly interesting to tie together the Higgs properties and a modified EW sector which can account for the increased \ww measurement, and we will report on this in~\cite{workinprogress}.

\section{Conclusions}
\label{conclusions} \setcounter{equation}{0} 

We have demonstrated that the current \ww cross section measurements favor the inclusion of a chargino contribution compared to the SM \ww prediction alone. This is not in conflict with any existing constraints, and improves the measurement at both LHC7 and LHC8 thus far.  
Given that the NLO SM prediction does not properly predict the data's normalization or shape in three independent measurements, it seems likely that there is either a significant gap in our understanding of the SM calculation {\em or} we are possibly getting a glimpse of new physics right at the EW scale. Either of these outcomes obviously carry profound consequences both theoretically and experimentally.  Any scenario involving new physics will have other experimental consequences which should be searched for while also improving the precision of the SM \ww measurement.  In the particular SUSY example that we put forward, continued multi-lepton studies should hopefully shed light on this scenario by the end of the LHC8 run. However, exclusion is intertwined with understanding the SM measurement and will become increasingly complicated if new physics is in this regime.  We have presented results for one benchmark SUSY scenario, however there are others that also can significantly improve the  \ww measurement.  There are additional possibilities beyond those presented here~\cite{workinprogress}, but clearly this letter should serve as a clarion call to both the theoretical and experimental communities to understand \ww production better.

\subsection*{Acknowledgements}

We would like to thank Antonio Delgado, Elliot Lipeles, Rohini Godbole, Sanjay Padhi, Marc-Andre Pleier, and James Wells for useful discussions. D.C. would also like to thank the CERN BSM-TH institute, where part of this research was conducted. The work of D.C. was supported in part by the National Science Foundation under Grant PHY-0969739. The work of P.J. was supported in part by the U.S.  National Science Foundation under grant NSF-PHY-0969510, the LHC Theory Initiative, Jonathan Bagger, PI. The work of P.M. was supported in part by NSF CAREER Award NSF-PHY-1056833.


\begin{thebibliography}{99}

\bibitem{wwatlas5}
``Measurement of the W+W- Production Cross Section in Proton-Proton Collisions at sqrt(s) = 7 TeV with the ATLAS Detector", ATLAS-CONF-2012-025

\bibitem{wwcms5}
``Measurement of WW production rate", CMS-PAS-SMP-12-005

\bibitem{wwcms8tev}
``Measurement of WW production rate", CMS-PAS-SMP-12-013


  \bibitem{charginoLEPbound}
J. Beringer et al. (Particle Data Group), Phys. Rev. D86, 010001 (2012)
  
  
  
  \bibitem{lightneutralino} 
  H.~K.~Dreiner, S.~Heinemeyer, O.~Kittel, U.~Langenfeld, A.~M.~Weber and G.~Weiglein,
  Eur.\ Phys.\ J.\ C {\bf 62}, 547 (2009)
  [arXiv:0901.3485 [hep-ph]].

\bibitem{atlastrinew} 
  G.~Aad {\it et al.}  [ATLAS Collaboration],
  arXiv:1208.3144 [hep-ex].



\bibitem{atlashbb} 
  G.~Aad {\it et al.}  [ATLAS Collaboration],
  arXiv:1207.0210 [hep-ex].
  
  
\bibitem{atlashobs} 
  G.~Aad {\it et al.}  [ATLAS Collaboration],
  [arXiv:1207.7214 [hep-ex]].

\bibitem{cmshobs} 
  S.~Chatrchyan {\it et al.}  [CMS Collaboration],
  Phys.\ Lett.\ B
  [arXiv:1207.7235 [hep-ex]].
  
  
  \bibitem{dieter} 
  B.~Feigl, H.~Rzehak and D.~Zeppenfeld,
  arXiv:1205.3468 [hep-ph].
  
  
  
 \bibitem{EwinoHiggs}
   M.~Carena, I.~Low and C.~E.~M.~Wagner,
  JHEP {\bf 1208}, 060 (2012)
  [arXiv:1206.1082 [hep-ph]];
   K.~Blum, R.~T.~D'Agnolo and J.~Fan,
  arXiv:1206.5303 [hep-ph];
    N.~Arkani-Hamed, K.~Blum, R.~T.~D'Agnolo and J.~Fan,
  arXiv:1207.4482 [hep-ph];
  
 \bibitem{EwinoHiggsReview}
   A.~Djouadi,
  Phys.\ Rept.\  {\bf 459}, 1 (2008)
  [hep-ph/0503173].
 
\bibitem{workinprogress}
D.~Curtin, P.~Jaiswal, and P.~Meade, {\em to appear}.

\bibitem{Kribs:2008hq}
G.~D. Kribs, A.~Martin, and T.~S. Roy, ``{Supersymmetry with a Chargino NLSP
  and Gravitino LSP},''
  \href{http://dx.doi.org/10.1088/1126-6708/2009/01/023}{{\em JHEP} {\bf 0901}
  (2009)  023}, \href{http://arxiv.org/abs/0807.4936}{{\tt arXiv:0807.4936
  [hep-ph]}}.



  \bibitem{CMStrilepton}
    S.~Chatrchyan {\it et al.}  [CMS Collaboration],
  arXiv:1204.5341 [hep-ex].
  
  
  \bibitem{ATLAStrilepton}
  G.~Aad {\it et al.}  [ATLAS Collaboration],
  [arXiv:1204.5638 [hep-ex]].
  
  
  
  
  \bibitem{prospino}
    W.~Beenakker, R.~Hopker and M.~Spira,
  hep-ph/9611232;
    W.~Beenakker, M.~Klasen, M.~Kramer, T.~Plehn, M.~Spira and P.~M.~Zerwas,
  Phys.\ Rev.\ Lett.\  {\bf 83}, 3780 (1999)
  [Erratum-ibid.\  {\bf 100}, 029901 (2008)]
  [hep-ph/9906298].
  
  
  
  \bibitem{pythia8}
  T.~Sjostrand, S.~Mrenna and P.~Z.~Skands,
  Comput.\ Phys.\ Commun.\  {\bf 178}, 852 (2008)
  [arXiv:0710.3820 [hep-ph]].
  
  
  
  \bibitem{pythia64}
    T.~Sjostrand, S.~Mrenna and P.~Z.~Skands,
  JHEP {\bf 0605}, 026 (2006)
  [hep-ph/0603175].
  
  
  \bibitem{fastjet}
    M.~Cacciari and G.~P.~Salam,
  Phys.\ Lett.\ B {\bf 641}, 57 (2006)
  [hep-ph/0512210];
    M.~Cacciari, G.~P.~Salam and G.~Soyez,
  Eur.\ Phys.\ J.\ C {\bf 72}, 1896 (2012)
  [arXiv:1111.6097 [hep-ph]].
  


  \bibitem{ewnlo} 
  A.~Bierweiler, T.~Kasprzik, H.~Kuhn and S.~Uccirati,
  arXiv:1208.3147 [hep-ph].
  
  
\bibitem{cmsmulti} 
  S.~Chatrchyan {\it et al.}  [CMS Collaboration],
  JHEP {\bf 1206}, 169 (2012)
  [arXiv:1204.5341 [hep-ex]].

\bibitem{atlasdinew} 
  G.~Aad {\it et al.}  [ATLAS Collaboration],
  arXiv:1208.2884 [hep-ex].
  
  

\bibitem{atlasdinewinc}
ATLAS Collaboration, ``Search for anomalous production of prompt like-sign lepton pairs with the ATLAS detector", ATLAS-CONF-2012-069.




\bibitem{ggmLHC} 
  Y.~Kats, P.~Meade, M.~Reece and D.~Shih,
  JHEP {\bf 1202}, 115 (2012)
  [arXiv:1110.6444 [hep-ph]].
  
\bibitem{LHCgaga}
{\bf ATLAS} Collaboration, ``{Search for SUSY and UED in Final States with
  Photons and Missing Transverse Energy with the ATLAS Detector},''
  ATL-PHYS-SLIDE-2011-523, \url{http://cdsweb.cern.ch/record/1380305};
{\bf CMS} Collaboration, ``{Search for supersymmetry with photons, jets and
  MET},'' CMS PAS SUS-11-009, \url{http://cdsweb.cern.ch/record/1377324}.

  
  
  
  
\bibitem{neutralinostevatron} 
P.~Meade, M.~Reece and D.~Shih,
  JHEP {\bf 1005}, 105 (2010)
  [arXiv:0911.4130 [hep-ph]].

\bibitem{atlastriold} 
  G.~Aad {\it et al.}  [ATLAS Collaboration],
  Phys.\ Rev.\ Lett.\  {\bf 108}, 261804 (2012)
  [arXiv:1204.5638 [hep-ex]].
  
  

\bibitem{atlaswwres} 
  G.~Aad {\it et al.}  [ATLAS Collaboration],
  arXiv:1208.2880 [hep-ex].
  
  

\bibitem{susyprimer}
  S.~P.~Martin,
  In *Kane, G.L. (ed.): Perspectives on supersymmetry II* 1-153
  [hep-ph/9709356].





\bibitem{micromegas}
  G.~Belanger, F.~Boudjema, A.~Pukhov and A.~Semenov,
  Comput.\ Phys.\ Commun.\  {\bf 176}, 367 (2007)
  [hep-ph/0607059].
  


\bibitem{atlasHww} 
  G.~Aad {\it et al.}  [ATLAS Collaboration],
  Phys.\ Rev.\ Lett.\  {\bf 108}, 111802 (2012)
  [arXiv:1112.2577 [hep-ex]].

\bibitem{cmsHww}
  S.~Chatrchyan {\it et al.}  [CMS Collaboration],
  Phys.\ Lett.\ B {\bf 710}, 91 (2012)
  [arXiv:1202.1489 [hep-ex]].
  
  

 \bibitem{ATLASgammagamma}
   [ATLAS Collaboration],
  arXiv:1202.1414 [hep-ex].
 
 \bibitem{CMSgammagamma}
   S.~Chatrchyan {\it et al.}  [CMS Collaboration],
  arXiv:1202.1487 [hep-ex].
   



\end{thebibliography}
\end{document}